\documentclass[graybox]{svmult}

\usepackage{mathptmx}       
\usepackage{helvet}         
\usepackage{courier}        
\usepackage{makeidx}         
\usepackage{graphicx}        
\usepackage{multicol}        

\makeindex             

\begin{document}

\title*{The Importance of the Strategy in \hskip 10mm Backward Orbits}
\titlerunning{Importance of Strategy in Backward Orbits}
\author{Carmen Pellicer-Lostao and Ricardo L\'opez-Ruiz}
\institute{Carmen Pellicer-Lostao \at BIFI, University of Zaragoza, \email{carmen.pellicer@bifi.es}
\and Ricardo L\'opez-Ruiz \at Dept. of Computer Science \& BIFI, University of Zaragoza, \email{rilopez@unizar.es}}
\maketitle

\abstract*{This work considers reversed evolution in dynamical systems. In particular, asymptotic behavior of chaotic systems, when their orbits evolve backwards in time. Reversed dynamics reveals important aspects of the trajectories, such as a new necessary parameter. This is the strategy through which one orbit reaches an original state in the past. As a result, it is found that backward orbits exhibit sensitivity to the strategy. This gives additional evidence about the unpredictability of the past.}

\abstract{This work considers reversed evolution in dynamical systems. In particular, asymptotic behavior of chaotic systems, when their orbits evolve backwards in time. Reversed dynamics reveals important aspects of the trajectories, such as a new necessary parameter. This is the strategy through which one orbit reaches an original state in the past. As a result, it is found that backward orbits exhibit sensitivity to the strategy. This gives additional evidence about the unpredictability of the past.}

\section{Introduction}
\label{sec:1}
Traditionally, the study of dynamical systems has been mostly concerned about forward evolution, considering long term behavior of the orbits in the future. As a consequence of these studies, chaos theory developed since the 60's gave to an end with the ideas about the possibility of predicting the future \cite{Laplace1825} in chaotic systems. These ideas became the base of the well known ``butterfly effect", which is the property of non-linear systems to have sensibility to initial conditions  \cite{Poincare1892,Lorenz1963}. Since then, dynamical systems have been widely used to model many kind of phenomena showing complex evolution and unpredictability in the distant future \cite{Pomeau1985}, \cite{Kyrtsou2005}, \cite{May2007}, \cite{Pellicer2010}.

Conversely, backward evolution of dynamical systems \cite{Bennet1962,Davis1990,Ingram2002} can also be of interest to model complex phenomena \cite{Mira1996,Medio2006,Kenedy2007,Gardini2009}. Such as for example, to be able to predict the origin of the evolution of a complex system given a known present state. This work travels into the past states of the dynamical systems and analyzes the asymptotic behavior of backward orbits. In particular, it will try to unravel some amazing properties of predictability of earliest states of a system when coming from an given state in the present.

\section{Backward Trajectories}
\label{sec:2}
An $N$ dimensional iterative dynamical system is given by a function $F:U \subseteq R^N \rightarrow U$ that maps a state into a future state. Time is considered to be a discrete variable and they are formulated as follows:

\begin{eqnarray}
X_{t+1} = F (X_t)
\label{eq:01}
\end{eqnarray}

where $t = 0,1,\ldots,n$ represents the temporal variable, $X_0,X_1,\ldots,X_n$ are the states of the system in different instants of time and $U$ is the region of the $N$-dimensional space $R^N$ where the system evolves, also referred as the phase space.

The consecutive iterates of the system from an initial point $X_0$ is called the \textbf{forward orbit} of $X_0$ under $F$. It is customary to express the sequence of iterates that represent the forward orbit as $\{F^i(X_0)\}_{i=0}^{\infty}$, that is fully expanded in the following equation.

\begin{eqnarray}
\{F^0(X_0) = X_0, F^1(X_1) = F(X_0),\ldots,F^{n+1}(X_0) = F(F^n(X_0)),\ldots \}
\label{eq:02}
\end{eqnarray}

If the function $F$ is invertible, we can also talk about the \textbf{backward orbit} of $X_0$ under $F$, described as $\{F^{-i}(X_0)\}_{i=0}^{\infty}$. Pairing the time variable with the space variable gives us the full view of the evolution of the dynamical system:

\begin{eqnarray}
\{\ldots,(t_{-n}, F^{-n}(X_0)),.., (t_{-1}, F^{-1}(X_0)),(t_0,X_0),(t_1, F(X_0)),.., (t_n, F^n(X_0)), \ldots\}
\label{eq:03}
\end{eqnarray}

Considering chaotic systems, chaos requires $F$ to be a non-linear function. Consequently, the inverse map $F^{-1}$ is typically a multi-valued function. This means that there are multiple ways to map a unique future state into the past. Then, it is always going to be necessary to define a strategy to map a state into a previous one.

To see that, observe for example Fig. ~\ref{fig1}. In this figure, a one dimensional iterative dynamical system, the Tent map, and its inverse functions are depicted. In Fig. ~\ref{fig1} (right), it can be seen that two different prior states $X_{-1}$ are obtained, when iterating backwards from an initial state $X_0$.

To produce a reversed orbit or backward orbit, it is necessary to select iteratively one of these two values as we to follow our trip into the past.  The selection of a different path at any step means, that it is necessary to choose a different \textbf{strategy of backward evolution}. This is going to produce a different backward orbit and presumably a different original state. Then, we can conclude that \textbf{backward evolution is  deterministic} only when the strategy is fully known. Also, a future state is not necessarily linked to any fixed past state. This is so, in the sense that for a given present state there are different options, that turn up to be possible prior states.

The above discussion is related in some way to Iterated Function System (IFS) \cite{Barnsley1985} formed by the collection of its inverse functions.

In fact, IFS provides a convenient framework to study this collection of functions. However we have to take into account that IFS perspective is quite different to reverse dynamics. Actually, this is basically a geometric perspective, considering these functions compressors of the phase space and global constructors of fractals.

In this work, it is our interest to explore these systems from a temporal perspective. Considering the temporal dimension, backwards trajectories expose a new evidence of non-determinism worth to be explored. As a consequence, our objective is to find out the relevance of the backwards strategy and to seek for its significance in reversed evolution. To study this, the mechanism of calculating the backward orbits is expressed formally in the following section.

\section{A new Parameter in Backward Dynamics}
\label{sec:3}

Let's say, $F$ is an non-invertible chaotic map, whose inverse map is a multi-valued function $F^{-1}$. Let us suppose that the values of the inverse map at a point $X_{-t}$ are a total of $b$ possible $X_{-(t+1)}$ points, denoted as:

\begin{equation}
 \{F^{-1}(X_{-t})\}^{\{b\}}
\label{eq:04}
\end{equation}

To take a step backwards in the evolution of the system, it is necessary to choose one of these $b$ values. Let us say that this decision may be called a \textbf{backward selection} and let us represent it as $s$. In a backward iteration of $n$ steps, it is necessary to make $n$ of such selections. Then, a series of selections can be call a \textbf{strategy} of length $n$ and it will be denoted as $S^n$ . Consequently, the strategy can be expressed as a vector that stores the decisions taken at every step:

\begin{equation}
S^n=\{s_i\}_{i=1}^{i=n}=\{s_1,s_2,..,s_n\}
\label{equ:05}
\end{equation}

Here $s_i$ describes the backward selection at instant $-i$.

Now, let us discuss how to code the values of a backward selection, $s_i$. Following eq. \ref{eq:04}, the set of possible pasts of $X_{-t}$ at instant $-i$ can expressed as:
\begin{equation}
\{F^{-i}(X_{-t})\}^{\{b\}} = \{X_{-t-i}\}^{\{b\}}
\label{eq:06}
\end{equation}

To make a backward selection at instant $-i$ is nothing but to pick one state out of this set. Let us say, that state $k$ is chosen, being $k={0,1,..,b-1}$, and this state is denoted as $X_{-t-i}^{(k)}$. Then the backward selection at instant $-i$ is coded as $s_i=k$.

Observe that with $b$ possible values and $n$ back steps, there will a total of $b^n$ possible back strategies $S^n$. Then let us express any of these possible backward strategies as $S^n_r$, where the value of $r$ is coded as follows:

\begin{equation}
 r=s_1*b^{n-1}+...+s_n = \sum_{i=1}^{i=n} s_i*b^{n-i}
\label{eq:07}
\end{equation}

From the discussion above, let us conclude that the calculus of a backward trajectory of length $n$ from a present point $X_0=P$ according to the strategy $S^n_r$ can be obtained applying the following iterative procedure:

\begin{equation}
 X_{-i}^{(k)} = \{F^{-i}(X_{0})\}^{(k)},\hskip 3mm      when \hskip 3mm k=s_i
\label{eq:08}
\end{equation}

Here $k=0,1,...,b-1$ and the iteration step is denoted by $i=0,1,...,n$. This means that to calculate a backward trajectory from $P$ given a specific strategy $S^n_r$, we need to set the initial point in the present $X_0=P$, then calculate n-times the composite inverse map $F^{-n}$ choosing at each iteration $-i$ one state out of all the possible $b$ past states. The selection $X_{-i}^{(k)}$ is given by the value of $s_i=k$, the strategy of backward evolution at step $-i$.

The strategy for traveling into the past appears here as a new parameter in the evolution of the dynamic system. This parameter rules the strategy through which one orbit reaches an original state in the past. Now, the point of interest is to consider the predictability of the past states in terms of this strategy. To do that, a practical example is considered in the following section. This illustrates the relevance of the backwards strategy in the dynamics of the system.

\section{The Tent map moving backwards}
\label{sec:4}
A particular chaotic map is considered in order to illustrate the previous discussions and measuring them in full detail. Simple examples make relevant concepts more obvious. Then, for simplicity we take the skew tent map, whose $F$ and $F^{-1}$ are given by the following equations:

\begin{eqnarray}
x_{n+1} & = &
\left\{\begin{array}{l}
x_n/\alpha,\hskip 2.2cm  0 \leq x_n \leq \alpha, \\
(x_n -1)/(\alpha -1) ,\hskip 0.5cm \alpha \leq x_n \leq 1,
\end{array} \right. \label{eq:09} \\
x_{n-1} & = &
\left\{\begin{array}{l}
\alpha x_n,\hskip 2.4cm 0 \leq x_n \leq 1, \\
(\alpha -1)x_n +1,\hskip 1.03cm 0 \leq x_n \leq 1,
\end{array}\right. \label{eq:10}
\end{eqnarray}

This map has a parameter of evolution $\alpha$, where $\alpha \in [0,1]$.

\begin{figure}[htb]
\begin{center}
\includegraphics[scale=1.1]{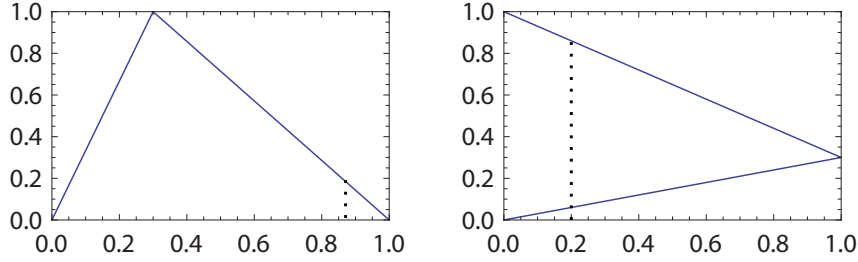}
\end{center}
\caption{(Left) Tent map and its inverse functions. (Right) Illustration of bi-valued past states in the inverse Tent map.}
\label{fig1}
\end{figure}

The Tent map and its inverse functions for the case of $\alpha=0.3$ are depicted in Fig. ~\ref{fig1}, right and left, respectively. The figures illustrate how a forward iteration calculates the value of $X_0= 0.2$ from a previous state $X_{-1}=0.86$. Continuing this route of evolution the map advances into the future. Conversely, backward iteration in the right panel of Fig. ~\ref{fig1} shows how a future state $X_0= 0.2$ produces two possible previous states $X_{-1}=0.86$ and $X_{-1}=0.06$.

In eqs. (\ref{eq:09}-\ref{eq:10}), it can be seen that two functions $F^{-1}[0,1]\rightarrow[0,1]$ are defined for the Tent map in the range $X_t \in [0,1]$. Then, the inverse map $F^{-1}(X_t)$ is a bi-valued function and there are $b=2$, two possible values of $X_{t-1}$ upon which we can make a single backward selection $s_{-1}$. These values are labeled $X_{t-1}^{(0)}$ and $X_{t-1}^{(1)}$.

Here the code of this selection means the following: when $k=0$ the point in the upper line of $F^{-1}$ is chosen, and when $k=1$ the lower line. Note that in this map when $X_t=1$, it occurs that $X_{t-1}^{(0)} = X_{t-1}^{(1)} = \alpha$. Then,  for a given subset $V \subset U$ and $X_t \in V$, it is possible that a different number of inverse options \cite{Mira1996} are found and so, different values of $b$ will exist for different $V$ depending on the number of precedents.

In a general form, it is possible to go $n$ steps backwards following  a given $S^n$ strategy. In this case $S^n$ is going to be a binary array of length $n$. The values $s_i$ of the strategy array $S^n$, are going to be either $0$ or $1$, depending on the selection of the $X_{-i}^{(0)}$ or $X_{-i}^{(1)}$, respectively.

Additionally, chaotic maps are dependent on the parameters of evolution. Then, backwards dynamics is dependent on these parameters. As we can see in equation eq.(\ref{eq:10}), the inverse functions obtained for the Tent map are dependent on $\alpha$ parameter . In the following sections this dependence is illustrated, combined with the strategy as well.

\subsection{Measuring parameters of Backward Evolution}
In this section we will consider how we represent a given strategy and the details of how this strategy rules the path to a given initial state in the past. To do that consider the case of moving from $X_0=0.2$, $n=5$ steps into the past. Then, one will find $b^n=2^5=32$ different possible backward strategies, and so $32$ different $X_{-5}$ past values giving rise to $X_0=0.2$ in the future.

\begin{figure}[htb]
\begin{center}
\includegraphics[scale=1.1]{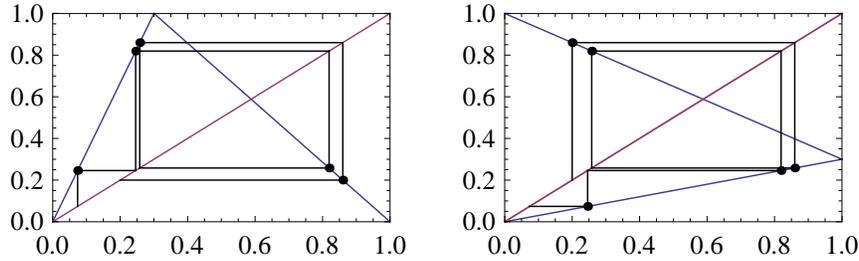}
\end{center}
\caption{(Left) Forward orbit $O_f$ and (Right) backward orbit $O_b$ to arrive $X_n=0.2$ in $5$ steps from $X_{n-5}=0.073746$.}
\label{fig2}
\end{figure}

Let's choose a strategy to travel into the past, such as for example $S^5_{11}=\{0,1,0,1,1\}$, where $r=11$ is calculated according to eq. (\ref{eq:07}). This particular $S^5_{11}$ means that we move backwards, choosing in the first step the upper branch of $F^{-1}$ in Fig. ~\ref{fig1} (Right), lower branch in the second step, upper in the third and so on. Table \ref{table1} and Fig. ~\ref{fig2} show the details of this particular example.

Table \ref{table1} shows the details of the particular backward selections taken at every step with strategy $S^5_{11}$. The resulting backward orbit is called $O_b$ and its values are $O_b=\{0.2, 0.86, 0.258, 0.8194, 0.24582, 0.073746\}$. As we can see in this table, this strategy lead the tent map to an initial state $X_{-5}=0.073746$. This table reveals the detail of every backward selection. At every step $-i$, two new possible values $\{F^{-i}(X_0)\}^{\{2\}}= \{ X_{-i}^{(0)}, X_{-i}^{(1)}\}$ are calculated.

The past state remains uncertain unless the strategy of backward selection is defined. Then, it is the value of $s_i$, the one that fixes the next step into the past, $X_{-i}^{(s_i)}$. The value of $X_{-i}^{(s_i)}$ is printed in bold in the table, to remark the selection taken at each step. Then, this value $X_{-i}^{(s_i)}$ produces two new possible past values $\{F^{-1}(X_{-i}^{(s_i)})\}^{\{2\}}$ in the next step and then, backwards iteration continues selecting one value of $\{X_{-i}\}^{\{2\}}$ according to $s_{i}$ until $i=n$ is reached.

\begin{table} [h]
\begin{center}
\begin{tabular}{|c|c|c|c|c|c|c|}
   \hline
  i & $0$ & $1$ & $2$ & $3$ & $4$ & $5$ \\
  \hline
  $X_{-i}^{(0)}$ & \textbf{0.2} & \textbf{0.86} & 0.398 & \textbf{0.8194} & 0.42624 & 0.827926 \\
  \hline
  $X_{-i}^{(1)}$ & - & 0.06 & \textbf{0.258} & 0.0774 & \textbf{0.2458} & \textbf{0.073746}\\
  \hline
  $s_{i}$ & - & 0 & 1 & 0 & 1 & 1 \\
  \hline
\end{tabular}
\caption{List of values for calculating backward trajectory $O_b$.}
\end{center}
\label{table1}
\end{table}

Fig. ~\ref{fig2} shows the graphics considered in this example, the forward orbit $O_f$ and the backward orbit $O_b$. The points of this orbits are $O_f=\{0.073746, 0.24582, 0.8194, \\0.258, 0.86, 0.2\}$ and $O_b=\{0.2, 0.86, 0.258, 0.8194, 0.24582, 0.073746\}$. In Fig. ~\ref{fig2}(Left), the forward orbit $O_f$ is obtained from its reversed associate $O_b$ Fig. ~\ref{fig2}(Right). Moving into the future from an initial state $X_{-5}=0.073746$ is a complete deterministic process. The Tent map evolves inevitably from $X_{-5}=0.073746$ to the future state $X_0=0.2$, following the determined orbit $O_f$.

In contrast, it is interesting to remark that, moving in reverse is a non deterministic process unless the strategy is fixed. Here Fig. ~\ref{fig2}(Right) shows in detail the points of $O_b$. $O_b$ is one of the $2^5=32$ possible backward orbits considered in this example. This particular trajectory $O_b$ is obtained moving from $X_0=0.2$ to $X_{-5}=0.073746$ according to a specific selected strategy, $S_{11}= \{0,1,0,1,1\}$.

\subsection{Deterministic Backward Evolution with a Strategy}
\begin{figure}[htb]
\begin{center}
\includegraphics[scale=.6]{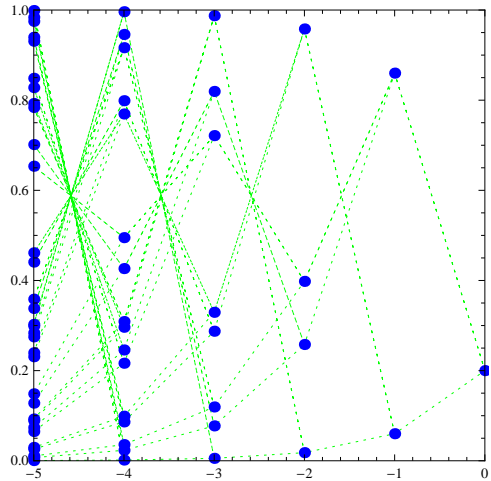}
\end{center}
\caption{Points of all possible $32$ backward orbits, that start from $X_0=0.2$, move $n=5$ steps into the past. A total of $2^{15}=32$ different initial past states are found for $X_{-5}$.}
\label{fig:03}
\end{figure}

It is observed that every strategy carries us to a particular different past state, while traveling through different branches of the inverse Tent map. From this, it's logical to think that if other branches are visited in the travel to the past, the initial state to which the system returns it is going to be different than $X_{-5}=0.073746$. To see this, let us move from $X_0=0.2$, $n=5$ steps into the past and compute all different backward origins $X_{-5}$ given by the $b^n=2^5=32$ different possible backward strategies $S_r^n$. Fig. ~\ref{fig:03} shows the backward computation of all possible values of $X_{-1},X_{-2},X_{-3}$ and $X_{-4}$ obtained at every step, up to reaching an earliest state $X_{-5}$. The $x$-axis presents the backward steps and the $y$-axis the different values of $X$ in the interval $[0,1]$ obtained at every step. The dot lines link the states obtained for every different possible strategy.

In Fig. ~\ref{fig:03}, it can be observed that moving $n=5$ steps backwards from an initial state $X_0=0.2$ is a non deterministic process. In fact, there are as may as $b^n=2^5=32$ strategies that lead the system to $32$ different initial past states. It can be seen that every strategy takes the system to a completely different $X_{-5}$ point in the past. The interested reader can easily recognize in this figure the particular backward orbit $O_b$ illustrated in Table \ref{table1} and Fig. ~\ref{fig2}(Right).

\begin{figure}[htb]
\begin{center}
\includegraphics[scale=1]{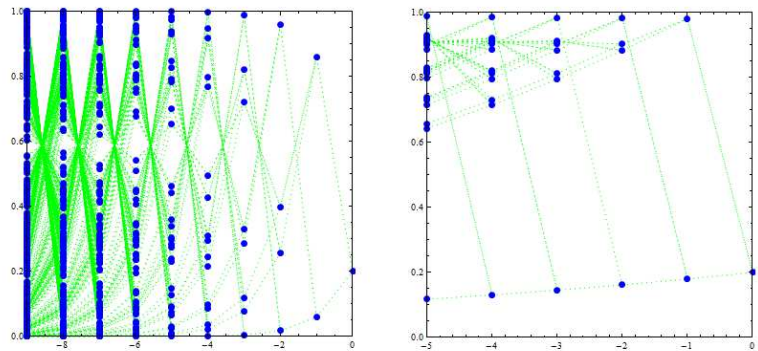}
\end{center}
\caption{(Left) Points of all possible $2^{10}$ backward orbits, that start from $X_0=0.2$ with $\alpha=0.3$, move $n=10$ steps into the past and arrive to $1024$ different initial past states, $X_{-10}$. (Right) Points of all possible $2^{5}$ backward orbits, that start from $X_0=0.2$ with $\alpha=0.9$, move $n=10$ steps into the past and arrive to $32$ different initial past states, $X_{-5}$.}
\label{fig:31}
\end{figure}

As a result, it can be said that \textbf{reversed dynamics is sensitive to the backwards strategy}. That situation is similar to the sensitivity to initial conditions observed in forward dynamics. Note that a small change in the trajectory, modifying just one $s_i$ will lead the system to a completely different original state. Also note that as we travel deeper into the past, the many more possible origins may appear and the origin of the system will be more difficult to predict, if the strategy is not recalled precisely. This can be explicitly seen in Fig ~\ref{fig:31}(Left) where we take the evolution of $n=10$ steps into the past instead of $n=5$ as in Fig. ~\ref{fig:03}. In Fig ~\ref{fig:31}(Left) there are $2^{10}$ possible backward orbits and the same number of possible past points $X_{-10}$. A small change in the strategy take us to a very different past state.

Therefore, similarly to the ``Butterfly effect" observed in forward dynamics, the sensitivity to the strategy tells us something important about the uncertainty of the past. It is impossible to predict the origin of a system unless the strategy is precisely known. It also can be said that, accurate data of a strategy may be unfeasible when the origin is remote and exceeds the physical capabilities of knowledge. This gives some evidence for the unpredictability of the past.

To illustrate dependence of backward dynamics to the other parameters of evolution, Fig ~\ref{fig:31}(Right) displays the $2^{5}$ possible backwards orbits that can be obtained moving $n=5$ steps backwards from an initial state $X_0=0.2$ when $\alpha=0.9$. This figure can be compared with Fig. ~\ref{fig:03} where $\alpha=0.3$ and compare the difference obtained in the orbits when the parameter $\alpha=0.9$.

\subsection{Sensitivity of Backward Evolution with the Strategy}

Finally, let us measure the sensitivity of past trajectories to the backward strategy. To do that, let us take the same case as before, traveling backwards $n$ steps into the past form $X_0=0.2$ with $F^{-1}$ of eq. (\ref{eq:09}-\ref{eq:10}) and $\alpha=0.3$. The initial state $X_{-n}$ is calculated for any of the $2^n$ possible strategies, that takes the inverse Tent map from $X_0=0.2$ to $X_{-n}$. As it was shown before, for every different strategy $S_r^n$ a different origin point $X_{-n}$ is produced. It can also be seen that as $n$ grows and the system travels deeper into the past, the values of $X_{-n}$ spread in a fractal way over the interval of the phase space $U=[0,1]$.

\begin{figure}[htb]
\begin{center}
\includegraphics[scale=0.7]{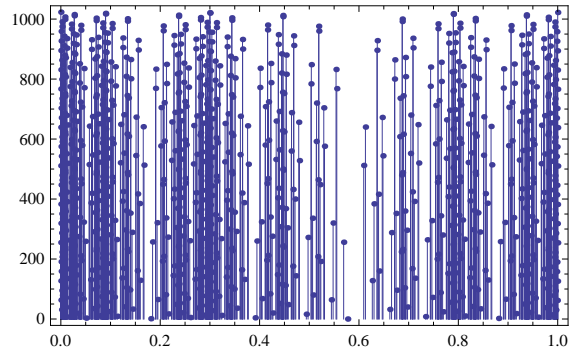}
\end{center}
\caption{Representations of $2^{10}=1024$ possible past points obtained for a trip of $n=10$ steps into the past from $X_0=0.2$ with $\alpha=0.3$.}
\label{fig4b}
\end{figure}

To illustrate these facts, the pairs $(X_{-n},r/2^{n})$ are plotted in Fig. ~\ref{fig4b}. Here $r$ is the number of the strategy $S^n_r$ that leads to the state $X_{-n}$ in the past. This number is normalized to one, taking $r/2^{n}$, in order to get a representation independent of $n$, the number of steps into the past. In Fig. ~\ref{fig4b}, a total of $n=10$ is considered and so, a total of $2^{10}=1024$ different strategies are depicted. The  $y$-axis represent the normalized value $r/2^{n}$ of the number of the strategy and the $x$-axis the past state $X_{-n}$ reached with strategy number $r$. This figure can be a more useful representation than Fig. ~\ref{fig:03} in order to show all the possible $X_{-n}$ states in a travel to a remoter past state.

As it is observed in this figure, the unpredictability of the original state can be appreciated graphically. This is due to the sensitivity to the backwards strategy. In particular, three important facts are observed. The first one is that the values of $X_{-n}$ spread in a fractal way over the phase space, the interval $U=[0,1]$, in accordance with IFS framework \cite{Barnsley1985}. The second is that when $n$ grows and the system travels deeper into the past, many more possible values of $X_{-n}$ arise. And the third one is that strategies differing just a single bit give very different initial states, that again spread in a fractal way over the phase space.

\begin{figure}[htb]
\begin{center}
\includegraphics[scale=1.1]{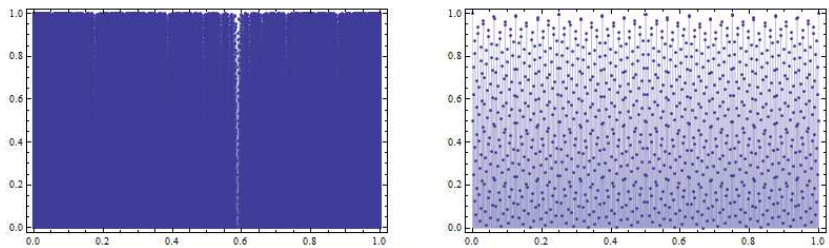}
\end{center}
\caption{(Left) Representations of $2^{18}=262144$ possible past points obtained for a trip of $n=18$ steps into the past from $X_0=0.2$ with $\alpha=0.3$. (Right) Representations of $2^{10}=1024$ possible past points obtained for a trip of $n=10$ steps into the past from $X_0=0.2$ with $\alpha=0.5$.}
\label{fig:041}
\end{figure}

This means that traveling into an initial state in the past requires recalling every decision in the strategy. If a single bit of the strategy is forgotten the system arrives to a different past origin. This can be explicitly seen in Fig ~\ref{fig:041}(Left) where we take the evolution of $n=18$ steps into the past instead of $n=10$ as in Fig. ~\ref{fig4b}. In Fig ~\ref{fig:041}(Left) there are $2^{18}$ possible backward orbits and the same number of possible past points $X_{-18}$. A small change in the number of the strategy take us to a very different past state.

These results give a new perspective for modeling the origins of complex systems. They offer a complementary point of view to the ``butterfly effect" observed in forward dynamics. The study of reversed dynamics reveals that it is impossible the discovery of the remote origin of complex phenomena. This is so, because this calculus exceeds the capabilities of knowledge, when the origin is far in the distant past. Put it in another words, for chaotic systems not only the far future, but also the remote past are unpredictable.

At this point let us remark that, the unpredictability of future phenomena has had great significance for applied sciences. The theory of complex systems has given new perspectives to sciences where chaotic behaviors have been observed like meteorology \cite{Lorenz1963}, economy \cite{Kyrtsou2005}, or others. In those sciences the discovery of the future has been granted as limited. One striking example can be the present economic and financial crisis, not predicted by anyone. Hence, the future is taken as uncertain and it is gradually enlightened at every forward step. Conversely, the acknowledge of the unpredictability of the past exposes a new perspective to applied sciences, that model the origin of complex phenomena. These sciences must consider the irony of this uncertainty and be aware that the discovery of past must be granted as limited. The past must be taken as uncertain, and it will only be gradually enlightened at every backward step we make.

To illustrate dependence of backward dynamics to the other parameters of evolution, Fig ~\ref{fig:041}(Right) displays the $2^{10}$ possible backwards orbits that can be obtained moving $n=10$ steps backwards from an initial state $X_0=0.2$ when $\alpha=0.5$. This figure can be compared with Fig. ~\ref{fig4b} where $\alpha=0.3$ and compare the difference of the fractal depicted in $x$-axis when the parameter $\alpha=0.5$.

\section{Conclusions}
\label{sec:4}

Reversed dynamics shows amazing mathematical aspects of the structure of the trajectories. Precisely, it can be said that it is possible to construct backward orbits and travel with reversed dynamics to the initial state of a non-linear system. This is done through a new parameter of dynamical evolution, the so called {\it the backward strategy}. Backwards dynamics demonstrates high sensitivity to this strategy. Hence, to calculate the earliest state of the system requires to recall precisely every step in the past history of the system. If a single bit of the strategy is forgotten or misunderstood, the system arrives to a completely different original state. In this sense and from an asymptotic perspective, it can be can concluded that the past is, in fact, unpredictable.

This can sound as a tautology but it could have some consequence in the studies of complex phenomena. In non-invertible dynamical systems not only the far future can be chaotic and unpredictable, but the remote past is also uncertain. Applied sciences that model the origin of the evolution of complex systems must be aware of that, just that the discovery of the past must be taken as limited.

In summary, this work portrays the relevance of the strategy in backwards orbits. Backwards dynamics turns out to be sensitive to the strategy and that makes it eligible as a new parameter of the evolution of dynamical systems. Considering that so, the strategy take us to the evidence of the unpredictability of the past.

%
%
%
%
%


\begin{thebibliography}{99.}

\bibitem{Barnsley1985}
Barnsley, M.F., Demko, S.: Iterated function systems and the global construction of fractals. The Proceedings of the Royal Society of London. A399, pp. 243-275 (1985)

\bibitem{Bennet1962}
Bennett, R.: On inverse Limit Sequences, Master's Thesis, University of Tennessee (1962)

\bibitem{Davis1990}
Davis, J. F.: Confluent mappings on $[0, 1]$ and inverse limits, Topology Proc. 15 pp. 1-9 (1990)

\bibitem{Gardini2009}
Gardini, L., Hommes, C., Tramontanac, F., de Vilderd, R.: Forward and backward dynamics in implicitly defined overlapping generations models. Journal of Economic Behavior and Organization \textbf{71} pp. 110–129 (2009)

\bibitem{Ingram2002}
Ingram, W.T.: Invariant sets and inverse limits. Topology and its Applications \textbf{126} pp. 393–408 (2002)

\bibitem{Kenedy2007}
Kennedy, J., Stockman, D. R., Yorke, J. A.: Inverse limits and an implicitly defined difference equation from economics. Topology and its Applications \textbf{154} pp. 2533–2552 (2007)

\bibitem{Kyrtsou2005}
Kyrtsou, C., Vorlow, C.: Complex dynamics in Macroeconomics: A novel approach. In: Diebolt, C., Kyrtsou, C. (eds.), New Trends in Macroeconomics,  pp. 223-245, Springer-Verlag, Berlin (2005)

\bibitem{Laplace1825}
Laplace, P.S.: Trait\'{e} de M\'{e}canique C\'{e}leste. Oeuvres compl\`{e}tes, Vol. 5, Gauthier-Villars, Paris (1825)

\bibitem{Lorenz1963}
Lorenz, E.N.: Deterministic non-periodic flow. J. Atmos. Sci., \textbf{20}, pp. 130-141 (1963)

\bibitem{May2007}
May, R.M., McLean, A.R.: Theoretical Ecology: Principles and Applications. Blackwell, Oxford (2007)

\bibitem{Medio2006}
Medio, A., Raines, B. E.: Inverse limit spaces arising from problems in economics. Topology and its Applications \textbf{153} pp. 3437-3449 (2006)

\bibitem{Mira1996}
Mira, C., Gardini, Barugola, L.A., Cathala, J.-C.: Chaotic Dynamics in Two-Dimensional Noninvertible Maps. series A, \textbf{20}, World Scientific Publishing, Singapore (1996)

\bibitem{Pellicer2010}
Pellicer-Lostao, C., L\'{o}pez-Ruiz, R.: A chaotic gas-like model for trading markets. Journal of Computer Science, \textbf{1}, pp. 24–32 (2010)

\bibitem{Poincare1892}
Poincar\'{e}, H.: Les m\'{e}thodes Nouvelles de la M\'{e}canique C\'{e}leste. 3 volumes, Gauthiers-Villars, Paris (1892)

\bibitem{Pomeau1985}
Berg\'{e}, P., Pomeau, Y., Vidal, C.: L'Ordre dans le Chaos. Hermann, Paris (1984)

\end{thebibliography}
\end{document}